\documentclass[fleqn,preprint,5p]{elsarticle}
\usepackage{graphicx}
\usepackage{graphics}
\usepackage[retainorgcmds]{IEEEtrantools}
\usepackage{color}
\usepackage{xspace} 
\usepackage{slashed}
\usepackage{amsmath}
\usepackage{amsfonts}
\usepackage{amssymb}

\newcommand{\rp}{r_{+}}

\begin{document}
\global\parskip 6pt

\author{Marc Casals}
\address{Centro Brasileiro de Pesquisas F\'isicas (CBPF), Rio de Janeiro, CEP 22290-180, Brazil.}
\address{School of Mathematical Sciences and Complex \& Adaptive Systems Laboratory, University College Dublin, Belfield, Dublin 4, Ireland.}
\ead{mcasals@cbpf.br, marc.casals@ucd.ie}

\author{Alessandro Fabbri}
\address{Centro Studi e Ricerche E. Fermi, Piazza del Viminale 1, 00184 Roma, Italy.}
\address{Dipartimento di Fisica dell'Universit\`a di Bologna and INFN sezione di Bologna, Via Irnerio 46, 40126 Bologna, Italy.}
\address{Laboratoire de Physique Th\'eorique, CNRS UMR 8627, B\^at. 210, Universit\'e Paris-Sud 11, Univ. Paris-Saclay, 91405 Orsay Cedex, France.}
\address{Departamento de F\'isica Te\'orica and IFIC, Universidad de Valencia-CSIC, C. Dr. Moliner 50, 46100 Burjassot, Spain.}
\ead{afabbri@ific.uv.es}

\author{Cristi\'an Mart\'{\i}nez}
\address{Centro de Estudios Cient\'{\i}ficos (CECs), Av. Arturo Prat 514, Valdivia, Chile.}
\ead{martinez@cecs.cl}

\author{Jorge Zanelli}
\address{Centro de Estudios Cient\'{\i}ficos (CECs), Av. Arturo Prat 514, Valdivia, Chile.}
\ead{z@cecs.cl}


\title{Quantum dress for a naked singularity}

\begin{abstract} 
\textbf{Abstract:} We investigate  semiclassical backreaction on a conical naked singularity space-time with a negative cosmological constant in $(2+1)$-dimensions. In particular, we calculate the renormalized quantum stress-energy tensor for a conformally coupled scalar field on such naked singularity space-time. We then obtain the backreacted metric via the semiclassical Einstein equations. We show that, in the regime where the semiclassical approximation can be trusted,  backreaction dresses the naked singularity  with an event horizon, thus enforcing (weak) cosmic censorship.
\end{abstract}


\date{\today}
\maketitle

\section{Introduction}  

Naked Singularities (NS) in gravitation theory are irksome: the curvature tensor and the energy density can `blow up'; the space-time fabric  may fail to resemble a smooth manifold and it may not be possible to continue geodesics past them; the laws of physics and standard features like causality may be violated \cite{Earman}. If  singularities are hidden behind an event horizon, however, one can safely ignore the problem because no causal signal can reach an outside observer from the troublesome region. It is  the spirit of Penrose's  Cosmic Censorship hypothesis that  NSs do not occur in nature \cite{Penrose}. In its weak version, this hypothesis essentially states that, generically, no `naked' (i.e., without an event horizon) space-time singularities  can form in Nature. NSs have been seen to form in some settings in $(3+1)$-dimensions, e.g.,~\cite{ESC,Gundlach}, although how `natural' and `generic' these settings are may be a matter of debate; in higher dimensions, NSs have been seen to form in~\cite{lehner2010black}.  However, in none of these works, quantum effects were taken into account. That naturally leads to the question of whether NSs are stable under quantum effects or else, for example, these effects lead to the formation of an event horizon.

Quantum effects on a curved background space-time, however, are notoriously difficult to calculate. 
One way to incorporate quantum effects is to include them in the energy momentum tensor and to solve the `semiclassical' Einstein equations (Eq.(\ref{semi}) below) for the backreaction  on the metric. The quantized stress-energy tensor for matter fields suffers from well-known ultraviolet divergences and so it must be appropriately renormalized (see, e.g.,~\cite{Birrell:Davies}). Such renormalization and obtention of the corresponding backreacted gravitational field, however, is very hard to perform in practice in $(3+1)$-dimensions unless the background space-time is highly-symmetric --such as pure de Sitter or pure anti de Sitter, (A)dS, space-times--, which is not the case for a black hole or NS space-times  in $(3+1)$-dimensions. On the other hand, in a space-time with one dimension less it is possible to make significant analytical progress while the results still yield an important insight into the physical processes that take place and into what one might expect there to happen in similar settings in $(3+1)$-dimensions.

In this paper we will investigate conical defects/excesses in ($2+1$)-AdS space-time. These are a particular class of NSs that do not seem to give rise to catastrophic phenomena. Like in an ordinary cone, the curvature singularity is a Dirac delta distribution at the tip. The source that produces this curvature can be identified with a point particle, which can also be understood as a removed point from the manifold \cite{DJT}. The geometry with the conical singularity is obtained by identification under a Killing vector in the universal covering of anti de Sitter space-time, CAdS$_3$, in complete analogy with the 2+1 (BTZ) black hole \cite{BTZ,BHTZ}. Since the identification does not change the local geometry, the conical singularity is a locally AdS space-time. 

The static circularly symmetric metric in Schwarzschild-like coordinates, $-\infty <t <\infty, 0 \leq r <\infty, 0\leq \theta \leq 2\pi \approx 0$, is given by
\begin{align} \label{staticBTZ}
ds^2=-\left(\frac{r^2}{\ell^2}-M\right)dt^2+\left(\frac{r^2}{\ell^2}-M\right)^{-1}dr^2+r^2d\theta^2,
\end{align}
where the mass $M$ is an integration constant and the  cosmological constant is given by $\Lambda = -\ell^{-2}$~\cite{BTZ}. This metric corresponds to a family of extrema of the vacuum Einstein-Hilbert action in $(2+1)$-dimensions. In three dimensions, black holes and conical singularities are just different parts of the spectrum of pure gravity, with black holes occupying the mass range $M>0$ and naked conical singularities corresponding to $0> M \neq -1$. The case  $M=-1$ corresponds to  $\text{AdS}_3$.

The naked  singularity is of a conical type at $r=0$, with deficit/excess angle $\Delta\equiv 2\pi (1-\sqrt{-M})$:  for $0 > M > -1$ there is an angular defect, while for $-1> M$ there is an angular excess. For $M\rightarrow 0^-$ the conical deficit approaches $2\pi$ and the NS undergoes a topological transition: the cone becomes a cylinder. On the other side of the transition there is a black hole of vanishing mass $M=0^+$.  As shown in \cite{mivskovic2009negative}, conical singularities can also carry angular momentum $J$, with $M\leq -|J|$. In the extreme case $M=-|J|$, these spinning particles, like the extreme black holes counterparts ($M=|J|$), are BPS states, admitting a supersymmetric extension and enjoying perturbative stability \cite{C-H}.

The identification vector $\xi$ in CAdS$_3$ that produces the black hole has norm $\xi\cdot \xi = r^2$. Thus, the region where $\xi$ is spacelike ($r^2>0$) is defined as the BTZ space-time. The region where the vector is timelike is excised in order to avoid the closed timelike curves  produced by the identification, generating a causal boundary at $r=0$. On the other hand, the conical singularity is produced by identifying with a rotation Killing vector $\eta \equiv \Delta \partial_\theta$, in AdS$_3$, where $\theta$ is the azimuthal angle and $\Delta$ is the conical deficit around $r=0$. This Killing vector  is spacelike everywhere and therefore does not produce closed timelike curves. However, the identification gives rise to a conical singularity at $r=0$, the fixed point of $\eta$. The opposite of a conical defect, an angular excess, is also a NS with a ``negative angular deficit", which is not produced by an identification, but by an insertion of an angular wedge.

These features make conical singularities in AdS$_3$ as acceptable as black holes. The question we wish to address, then, is, what happens in the geometry of a conical singularity when one includes vacuum fluctuations of some matter field: does the conical defect of the NS grow? what is the fate of the singularity? In this paper we  investigate precisely this issue on a non-rotating, naked conical singularity space-time in $(2+1)$-dimensions with a negative cosmological constant and find that quantum effects create an event horizon surrounding a curvature singularity.

This paper is organized as follows. 
In Sec.\ref{sec:RSET}, we calculate the renormalized expectation value of the stress-energy tensor for  a  scalar field in a NS space-time after
reviewing the corresponding literature result in a black hole space-time.
In Sec.\ref{sec:NS} we analytically calculate the quantum-backreacted metric.
We finish in Sec.\ref{sec:discussion} with a discussion of our results.
We use units $c=1$, $G=1/8$ throughout.

\section{Renormalized Stress-energy Tensor}  \label{sec:RSET}

In \cite{martinez1997back} it was shown that quantum fluctuations of a scalar field with periodic boundary conditions around a black hole make its horizon radius grow\footnote{This result was extended to non-conformal coupling for the massless black hole in~\cite{binosi1999quantum} and 
to the four-dimensional planar massless black hole metric, in the conformal case, in~\cite{caldarelli1998quantum}.}. 
Hence, a black hole will remain a black hole if the quantum fluctuations are included. Here we explore the effect of quantum fluctuations on a conical singularity: does the conical singularity remain naked, or do the quantum corrections dress this singularity with an event horizon? Our analysis shows that the latter is the case. A similar question in flat (zero cosmological constant) $(2+1)$-dimensional space-time was raised by Souradeep and Sahni~\cite{Souradeep-Sahni} and by Soleng~\cite{soleng1993inverse}, who showed that quantum effects on a conical singularity in flat space turn it into a  $(2+1)$-dimensional `Schwarzschild-like' space-time with gravitational attraction. 
In flat 2+1 space-time, an analogous question was also addressed in the context of an accelerated C-metric in~\cite{anber2008ads4}.

In order to address the  above question, we consider the semiclassical Einstein equations
\begin{equation} \label{semi}
G_{\mu \nu} - \ell^{-2} g_{\mu \nu}= \kappa \left\langle \hat T_{\mu  \nu} \right\rangle_{\text{ren}} \,,
\end{equation}
where $G_{\mu \nu}$ is the Einstein tensor for the metric $g_{\mu \nu}$ and $\kappa=8\pi G$. These equations determine the perturbed metric via the renormalized expectation value of the stress-energy tensor (RSET), $ \left\langle \hat T_{\mu  \nu} \right\rangle_{\text{ren}}$, of the matter field in some quantum state. 
We consider as quantum source a conformally coupled  scalar field without a mass parameter, whose (unrenormalized) expectation value of the stress-energy tensor is given by (\cite{Birrell:Davies,steif1994quantum}):
\begin{align}\label{T-munu}
\langle \hat T_{\mu\nu}(x)\rangle =
&
\lim_{x' \rightarrow x}\frac{\hbar}{4}\left[ 3\nabla^x_\mu \nabla^{x'}_\nu - g_{\mu \nu} g^{\alpha \beta} \nabla^x_\alpha \nabla^{x'}_\beta 
\right.\nonumber \\ &\left.
- \nabla^x_\mu \nabla^x_\nu - \frac{1}{4\ell^2}g_{\mu \nu} \right] G(x,x'),
\end{align}
where $x$ and $x'$ are space-time points. Here, $G(x,x')$ is Hadamard's elementary two-point function, i.e., the anticommutator 
$\langle \{\hat \Phi (x),\hat \Phi (x')\}\rangle$,
where $\hat\Phi(x)$ is the quantum scalar field. The quantum state of the field where the expectation values of the stress-energy tensor and of the two-point function are evaluated is determined by imposing boundary conditions on the solutions of the field equations.
 In the present analysis, we choose for  the two-point function  $G(x,x')$ for the scalar field to satisfy `transparent' boundary conditions \cite{martinez1997back}.
Imposing transparent boundary conditions corresponds to quantizing the scalar field using modes which are
smooth on the entire Einstein static universe~\cite{avis1978quantum,Lifschytz:1993eb}.

We first review the calculation of the RSET existing in the literature in  the case of the black hole and afterwards
we derive our new results in the case of the NS.

\subsection{Black hole case ($M>0$)}   

In the BTZ black hole case, the  RSET in Eq.(\ref{T-munu}) when the scalar field satisfies `transparent' boundary conditions  takes the form (\cite{martinez1997back}, \cite{steif1994quantum}):
\begin{equation}\label{eq:RSET BH}
\kappa \langle \hat T^{\mu}{}_{\nu}\rangle_{\text{ren}} =\frac{l_P}{r^3}F_{BH}(M)\text{diag}(1,1,-2),
\end{equation}
in $\{t,r,\theta\}$ coordinates, where $l_P= \hbar G$ is the Planck length and $F_{BH}(M)$ is a function that we give in Eq.(\ref{F-btz}) below. The two-point function in the BTZ black hole case can be calculated via the method of images from the two-point function in $\text{AdS}_3$, taking advantage of the fact that the black-hole manifold $\mathcal{M}_{BTZ}$ is obtained by an identification of the universal covering space $\text{CAdS}_3$ of  $\text{AdS}_3$ under one of its isometries. Explicitly, $\mathcal{M}_{BTZ} \approx \text{AdS}_3/H_\xi$, where $H_\xi$ is the discrete group obtained by  applying a Lorentz boost $\xi$ \cite{BHTZ}.
The method of images then leads to an expression for the two-point function in $\mathcal{M}_{BTZ}$ in terms of the two-point function in $\text{CAdS}_3$ for a scalar field $\Phi$ with periodic boundary conditions $\Phi(\theta)=\Phi(\theta+2\pi)$ (where the coordinates other than $\theta$ are suppressed):
\begin{equation} \label{Gbh}
G_{BH}(x,x')=\sum_{n\in\mathbb{Z}}G_{CAdS_3}(x, H^n_\xi x'),
\end{equation}
The two-point function in $\text{CAdS}_3$  is (e.g., \cite{steif1994quantum}, \cite{shiraishi1994quantum},\cite{carlip19952+},\cite{Decanini:Folacci:2005a}),
\begin{equation} \label{GcAdS}
G_{CAdS_3}(x, x')=\frac{1}{4\pi}\frac{1}{|x-x'|},
\end{equation}
where $|x-x'|=\sqrt{(x-x')^a(x-x')_a}$ is the geodesic distance between $x$ and $x'$ in the embedding space $\mathbb{R}^{2,2}$. Representing the points in $\text{CAdS}_3$ as embedded in flat $\mathbb{R}^{2,2}$ as $\left(x^a\right)=\left(T_1,X_1,T_2,X_2\right)^T$, where $T_{1,2}$ are time coordinates and $X_{1,2}$ spatial ones, the identification takes the form
\begin{equation}\label{H}
H_\xi=
 \left( \begin{array}{cccc}
\cosh(2\pi\sqrt{M})  & \sinh(2\pi\sqrt{M})  & 0 & 0 \\
\sinh(2\pi\sqrt{M})  & \cosh(2\pi\sqrt{M})  & 0 & 0 \\
0 & 0 & 1 & 0 \\
0 & 0 & 0 & 1
 \end{array} \right).
\end{equation}
By noting that the space-time is locally AdS, it follows that the two-point function in Eq.(\ref{Gbh}) is renormalized  by subtracting the two-point function in $\text{CAdS}_3$, Eq.(\ref{GcAdS}); this amounts to merely subtracting the $n=0$ term  in Eq.(\ref{Gbh}) \cite{PhysRevD.40.948,Souradeep-Sahni,shiraishi1994quantum}. Finally, combining Eqs.(\ref{Gbh}), (\ref{GcAdS}) and (\ref{H}),  the RSET for a black hole is obtained as in Eq.(\ref{eq:RSET BH}), where \cite{steif1994quantum}
\begin{equation} \label{F-btz}
F_{BH}(M)= \frac{M^{3/2}}{2\sqrt{2}} \sum_{n=1}^{\infty}\frac{\cosh\left(2n\pi\sqrt{M}\right)+3}{\left(\cosh\left(2n\pi\sqrt{M}\right)-1\right)^{3/2}}.
\end{equation}

\subsection{Naked singularity case ($M<0$)}\label{sec:NS}  
It is possible to find the expressions corresponding to Eqs.(\ref{eq:RSET BH}), (\ref{Gbh}), (\ref{H}) and (\ref{F-btz}) for the  NS space-time
with $-1<M<0$, because the conical defect space-time is also obtained by an identification in the AdS$_3$ geometry given by (\ref{staticBTZ}) with $M=-1$. The only difference is that the identification Killing vector $\eta$ is not along a boost, but a spatial rotation that generates the angular deficit. The coordinate transformation between the coordinates in $\mathbb{R}^{2,2}$ and
$\bar t\equiv \sqrt{-M} t$,  $\bar r\equiv  r/\sqrt{-M}$ and  $\bar \theta\equiv \sqrt{-M} \theta$
in the NS space-time is \cite{mivskovic2009negative}:
\begin{align}\label{eq:coord transf NS}
&
T_1=\sqrt{\bar r^2+\ell^2}\cos \left(\bar t/\ell\right),\quad
X_1=\bar r\cos \bar \theta,
\nonumber \\ &
T_2=\sqrt{\bar r^2+\ell^2}\sin \left(\bar t/\ell\right),\quad
X_2=\bar r\sin \bar \theta.
\end{align}
Note that the angle $\theta$ is now on the $X_1-X_2$ plane (as opposed to on the $X_1-T_1$ plane in the BTZ black hole case). In these barred coordinates, Eq. (\ref{staticBTZ}) becomes
\begin{align} \label{eq:metric NS}
ds^2=-\left(\frac{\bar r^2}{\ell^2}+1\right)d\bar t^2+\left(\frac{\bar r^2}{\ell^2}+1\right)^{-1}d\bar r^2+\bar r^2d\bar\theta^2.
\end{align}
As $\theta$ has a period of  $2\pi$, then $\bar{\theta}$ has a period of $2\pi  \sqrt{-M}$, which clearly represents a conical singularity. This NS is obtained from $\text{CAdS}_3$ under the identification of $\bar\theta$ and $\bar\theta+2\pi \sqrt{-M}$, which is obtained by the matrix
\begin{equation}\label{eq:bar H}
 H_{\eta} =
 \left( \begin{array}{cccc}
\cos(2\pi\sqrt{-M})  & \sin(2\pi\sqrt{-M})  & 0 & 0 \\
-\sin(2\pi\sqrt{-M})  & \cos(2\pi\sqrt{-M})  & 0 & 0 \\
0 & 0 & 1 & 0 \\
0 & 0 & 0 & 1
 \end{array} \right), 
\end{equation}
in coordinates $\left(x^a\right)=\left(X_1,X_2,T_1,T_2\right)^T$, where the angular deficit is $\Delta\equiv 2\pi (1-\sqrt{-M}) \in(0,2\pi)$. The resulting NS metric is given by Eq.(\ref{staticBTZ}), where $-1<M<0$.

The two-point function for a conformally-coupled and massless scalar field satisfying `transparent' boundary conditions in NS can be given by the method of images (cf.  \cite{Souradeep-Sahni,martinez1997back,steif1994quantum}) in the case of  an angular deficit $\Delta=2\pi (1-1/N)$ (i.e., $M=-1/N^2$), with $N=1, 2, 3, ...$ In this case, the two-point function is given by
\begin{equation}\label{Green-NS}
G_{NS}(x,x')=\sum_{n=0}^{N-1}G_{CAdS_3}(x, H^n_{\eta} x')= \frac{1}{4\pi} \sum_{n=0}^{N-1} \frac{1}{|x- H^n_{\eta} x'|}.
\end{equation}

In the conical geometry this sum contains a finite number of terms because for two points on the surface of a cone there is a finite number $N$ of geodesics connecting them.\footnote{This can be easily understood considering a cone in $\mathbb{R}^2$: For $0<\Delta<\pi$  there is a unique geodesic connecting two points ($N=1$). For $\Delta=\pi$, $N=2$; for $\Delta = 4\pi/3$, $N=3$; in general, for $\Delta = 2\pi (k-1)/k$, $N=k$. Finally, for $\Delta \rightarrow 2\pi$ the cone approaches a cylinder and the number of geodesics becomes infinite \cite{Matschull}.} 

For $\Delta \neq 2\pi (1-1/N)$, the method of images does not apply and the two-point function must be computed as a sum over the field modes on the cone. The construction in this case follows the procedure of \cite{Souradeep-Sahni}, where instead of the Bessel function that appears in the expression for the two-point function in flat conical space, one finds associated Legendre functions with continuous degree and order \cite{Hobson}. In this way, the two-point function is found  to be continuous in $M$ and coincides with the expression in Eq.(\ref{Green-NS}) when $M=-1/N^2$ (i.e., $\Delta = 2\pi (1-1/N)$). 
In  the case of  angular excesses ($M < -1$, $\Delta<0$), Eqs.(\ref{eq:coord transf NS}), (\ref{eq:metric NS}) and (\ref{eq:bar H}) also apply,
there is a unique geodesic joining two space-time points and the method of images fails to be adequate as well.

The rationale is the same as in electrostatics: the method of images for a point charge between two conducting plates forming an angle $\theta=2\pi/p$ produces a finite number of images for rational $p$, otherwise the required images are infinitely many and densely distributed. This does not happen in the black hole case, for the same reason that the method of images for two parallel plates does not depend on the separation between the conductors, and also requires an infinite countable number of images, as in Eq.(\ref{Gbh}).

In analogy with the black hole case, the two-point function (\ref{Green-NS}) is to be renormalized by dropping the $n=0$ term. The RSET in this case is then given by
\begin{equation}\label{eq:RSET BH'}
\kappa \langle \hat T^{ \mu}{}_{ \nu}\rangle_{\text{ren}}  =\frac{l_P}{ r^3}  F_{NS}(M)\text{diag}(1,1,-2),
\end{equation}
 in $\{t,r,\theta\}$ coordinates, with
\begin{equation}\label{eq:FNS}
 F_{NS}(M)= \frac{(-M)^{3/2}}{2\sqrt{2}} \sum_{n=1}^{N-1} \frac{\cos (2n\pi\sqrt{-M} )+3}{\left(1-\cos (2n\pi\sqrt{-M})\right)^{3/2}}.
\end{equation}

We note that $F_{NS}(M)$ can be obtained from $F_{BH}(M)$ by analytic continuation, except for the important difference that Eq.(\ref{eq:FNS}) possesses a finite sum (as opposed to Eq.(\ref{F-btz})) and consequently $F_{NS}$ is manifestly finite. We also note that both  $F_{NS}(M)$  and  $F_{BH}(M)$ are positive definite within their respective mass ranges. 

The value of $F_{NS}(0)$ may be easily obtained  by taking the limit $M=-1/N^2 \to 0^-$ (i.e., $N\to \infty$ and $\Delta \to 2\pi$) in Eq.(\ref{eq:FNS}) and applying L'H\^opital rule. One readily finds
\begin{equation} \label{FNS(0-)}
F_{NS}(0^-)=\frac{\zeta(3)}{2\pi^3},
\end{equation}
where $\zeta$ is the Riemann zeta function. This value is the same as the limit $M\to 0^+$ in Eq.(\ref{F-btz}).

\section{Backreacted Metric} \label{sec:bacreaction}
In \cite{martinez1997back}, the backreaction on the geometry produced by a correction of the form 
\begin{equation}
\kappa \langle \hat T^{\mu}{}_{\nu}\rangle_{\text{ren}}  =\frac{A(M)}{r^3}\text{diag}(1,1,-2),
\end{equation}
as in  Eq.(\ref{eq:RSET BH'}), was computed, showing that the metric takes the form of the exact solution in the presence of a conformally coupled scalar field. Following the same steps, in our case we find
\begin{align} \label{eq:metric NS unbar}
ds^2=
&-\left(\frac{r^2}{\ell^{2}}-M-\frac{2l_P F_{NS}(M)}{r}\right) dt^2
\nonumber \\ &
 + \left(\frac{r^2}{\ell^{2}}-M-\frac{2l_P F_{NS}(M)}{r}\right)^{-1} dr^2 + r^2 d \theta^2.
\end{align}
This metric now has an event horizon since the equation
\begin{equation}\label{eqh}
 \frac{r^2}{\ell^{2}}-M=\frac{2l_P F_{NS}(M)}{r}
\end{equation}
has one positive root for any value of $M$ since $F_{NS}>0$ (see Fig.1).\footnote{The other two roots are complex.} 
\begin{figure}[h]
\centering 
\includegraphics[angle=0,width=0.40\textwidth]{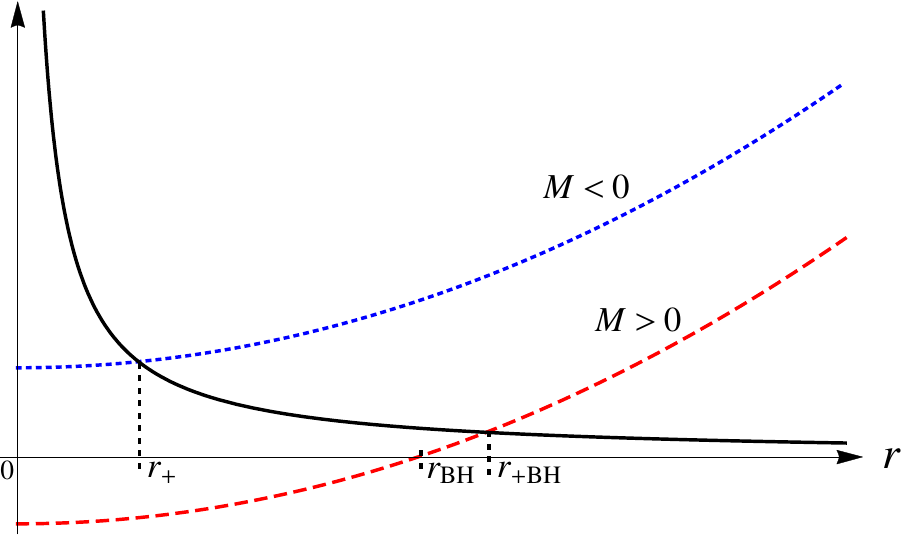}
\caption{Schematic description of the roots of Eq.(\ref{eqh}). The  black (straight) curve is the right hand side of Eq.(\ref{eqh}) and the blue (dotted) and red (dashed) curves are the left hand side of Eq.(\ref{eqh}) for, respectively, $M<0$ and $M>0$. The radius $\rp$ where the black  and blue curves meet corresponds to the  horizon of the backreacted NS geometry;
the radius $r_{+ \text{BH}}$ where the black and red curves meet corresponds to the  horizon of the backreacted black hole geometry;
 the radius $r_{\text{BH}}$ where the red curve crosses the horizontal axis corresponds to the horizon of the classical BTZ black hole.}
\label{fig:fig1} 
\end{figure}
For $M<0$ the horizon radius is given by
\begin{align} \label{exact}
&\rp=\frac{b}{3}+\frac{M \ell^2}{b}, \quad \mbox{with}
\end{align}
\begin{equation*}
b = \left (27 F_{NS}(M) \ell^2 l_P+ 3\ell^2\sqrt{81 F_{NS}(M)^2  l_P^2-3M ^3 \ell^2}\right)^{\frac{1}{3}} >0. 
\end{equation*}
The above metric has a curvature singularity at $r=0$, as shown by the Kretschmann scalar
\begin{equation}\label{eq:Kretschmann}
K= R_{\alpha\beta\gamma\delta}R^{\alpha\beta\gamma\delta}  =12 \left(\frac{1}{\ell^4} + \frac{2    (l_P F_{NS}(M))^2}{r^6}\right).
\end{equation}
Thus, similarly to the backreacted black hole \cite{martinez1997back}, a curvature singularity at $r=0$ is  generated. However, this singularity would lie inside a horizon with a radius that vanishes for $l_P \to 0$.  For $r \gg \rp$, the geometry approaches  AdS space-time.

It is interesting to consider two different limits. The first one consists of taking $l_P\to 0$ with finite $M$. From Eq.(\ref{exact}) one obtains
\begin{equation} \label{h1}
\rp = \frac{2l_P F_{NS}(M)}{-M}+O(l_P^3/\ell^3).
\end{equation}
Note that in the classical limit, $l_P \rightarrow 0$, $\rp$ goes to zero, while the two complex roots of Eq.(\ref{eqh}) approach $\pm i \ell\sqrt{-M}$. This means that the positive real root is a purely quantum effect that enforces cosmic censorship. No matter how large the conical defect is, quantum corrections of the vacuum would dress up the naked singularity.

For the expansion on the right hand side of Eq.(\ref{h1}) to make sense one cannot take $M\rightarrow 0^-$. This limit, with small but finite $l_P$, can be explored from Eq.(\ref{eqh}), which gives 
\begin{equation} \label{h2}
\rp = \frac{(l_P  \ell^2\zeta(3))^{\frac{1}{3}}}{\pi} +O(M).
\end{equation}

\section{Discussion}\label{sec:discussion}
The backreacted metric Eq.(\ref{eq:metric NS unbar}) shows a horizon forming at a finite radius $\rp$. However, for finite $M$ the horizon radius is of order $l_P$. This indicates that, in order to resolve a region of  size $r \sim \rp$, it would be required to go into the strong quantum gravity regime, casting doubt on the meaning of the classical description of space-time. Classical notions like metric, distance and curvature are meaningful for coarser resolutions, which can be applied for large distances and far away from the horizon. Thus, for finite $M$ the classical theory breaks down for $r \to 0$ and there is no ground to claim that a space-time singularity --naked or otherwise-- exists, because we have no theoretical framework to describe the space-time for $r \lesssim \rp$.

For small $M$, on the other hand, Eq.(\ref{h2}) implies $\rp/\ell \sim (l_P/\ell)^{1/3}$, which means $\rp^3 \sim \ell^2 l_P$ and, therefore, in the semiclassical approximation, $\rp \gg l_P$. This gives support to the interpretation of $\rp$ as a classical notion so that the claim that a horizon forms may be trusted. Moreover, as mentioned earlier, in the limit $M\to 0^+$ Eq.(\ref{F-btz}) coincides with Eq.(\ref{FNS(0-)}),
\begin{equation} \label{FNS-BH(0)}
F_{BH}(0^+)=\frac{\zeta(3)}{2\pi^3}=F_{NS}(0^-)\, .
\end{equation}
Hence,  one can expect that quantum corrections on a conical singularity with a deficit angle approaching $2\pi$, turn it into a state indistinguishable from a small mass black hole.

The Kretschmann invariant Eq.(\ref{eq:Kretschmann}) shows a strong curvature singularity forming at $r=0$, which seems to make matters worse than in the original conical singularity. However, as we saw in the previous discussion, for finite $M$ the semiclassical approximation is inadequate for describing the central region of the space-time. In the small $M$ approximation, however, the semiclassical approximation can be trusted and the central singularity --if any-- would be hidden by a horizon. In this latter case, substituting Eq.(\ref{h2}) in Eq.(\ref{eq:Kretschmann}) gives $K(\rp) \sim 18/\ell^4$, which is 50\% greater than the classical value.

In \cite{efk}, brane-localised black holes in $AdS_4$ were interpreted, via AdS/CFT, as quantum-corrected BTZ black holes. In the classical NS regime, it was noted that no calculation of the stress energy tensor of a conformal field in conical ($M<0$)  $AdS_3$ space-time was found in the literature to compare with, nor of its backreaction. Our paper fills this gap in the literature and confirms, in agreement with the analysis of \cite{efk}, that quantum effects can censor singularities.

The analysis presented here can be extended to include angular momentum $M<-|J|$. Among the spinning cases, the extremal ones, $M=-|J|$, are particularly interesting because they are BPS configurations and not expected to receive quantum corrections \cite{mivskovic2009negative}. We leave these questions for future work. Another interesting case to study is the conformally coupled scalar field system in 2+1 gravity, where a soliton with negative mass exists, which represents a non-trivial vacuum for hairy black hole sector \cite{CMT}.

\section*{Acknowledgments}
We would like to thank F. Canfora, A. C. Ottewill, P. Taylor, D. Tempo, R. Troncoso and T. Zojer for valuable comments and observations. This work has been partially funded throught grants 1130658, 1140155 and 1161311 from Fondecyt. The Centro de Estudios Cient\'{\i}ficos (CECs) is funded by the Chilean Government through the Centers of Excellence Base Financing Program of Conicyt.
M.C. acknowledges partial financial support by CNPq (Brazil), process number 308556/2014-3. A.F. acknowledges partial financial support from the Spanish MINECO through the grant 
FIS2014-57387-C3-1-P and the Severo Ochoa Excellence Center Project SEV-2014-0398.

\bibliographystyle{apsrev}

\end{document}